%% file: main.tex
\documentclass{article}

\PassOptionsToPackage{numbers, compress}{natbib}

\usepackage[dblblindworkshop, final]{neurips_2026}
\workshoptitle{Self-Evolving Diversity-Driven Search for Robust AI Systems}
\usepackage[utf8]{inputenc}
\usepackage[T1]{fontenc}
\usepackage{hyperref}
\usepackage{url}
\usepackage{booktabs}
\usepackage{tabularx}
\usepackage{amsmath}
\usepackage{amssymb}
\usepackage{amsfonts}
\usepackage{graphicx}
\usepackage{float}
\usepackage{xcolor}
\usepackage{algorithm}
\usepackage{algpseudocode}
\usepackage{microtype}
\usepackage[most]{tcolorbox}
\usepackage{caption}
\usepackage{multirow}

\definecolor{reqbg}{HTML}{E2DED6}
\newtcolorbox{reqbox}{
  enhanced, sharp corners,
  colback=reqbg, colframe=reqbg, boxrule=0pt,
  left=3pt, right=3pt, top=2pt, bottom=3pt,
  before skip=4pt, after skip=6pt,
  fontupper=\small
}

\title{Beyond User-Side Defenses:\\Image Protection in Open-Weight Diffusion Models}

\author{%
  Jihye Kang \qquad Jiwon Park \qquad Namhyuk Ahn \\
  Department of Electrical and Computer Engineering, Inha University \\
}

\begin{document}

\maketitle

\input{sections/0_abstract}
\input{sections/1_introduction}
\input{sections/related_work}
\input{sections/2_requirements}
\input{sections/3_method}
\input{sections/4_experiments}
\input{sections/5_conclusion}

\clearpage
\noindent\textbf{Acknowledgments.}
This work was supported by the National Research Foundation of Korea (NRF) grant funded by the Korea government (MSIT) (RS-2026-25603519).

{
\small
\bibliographystyle{unsrt}
\bibliography{main}
}

\clearpage
\appendix
\raggedbottom
\input{sections/appendix}

\normalsize

\end{document}

%% file: sections/0_abstract.tex
\begin{abstract}
Model providers can support image protection, but open-weight deployment introduces practical constraints.
We propose four requirements: protection effectiveness, clean fidelity, compatibility, and tamper resistance.
We incorporate long-horizon attack simulations to strengthen resistance to protection removal.
Experiments with a single protection tool show improved resistance to re-alignment at a cost to clean fidelity, motivating further research toward practical model-side protection.
\end{abstract}

%% file: sections/1_introduction.tex
\section{Introduction}

Generative models can imitate artistic styles~\cite{dreamstyler,zhang2024finestyle}, reproduce personal likenesses~\cite{dreambooth}, and manipulate existing images~\cite{yu2025anyedit}, raising concerns about copyright protection and the misuse of personal images~\cite{glaze}.
Image protection methods seek to limit such unauthorized uses by adding small perturbations that interfere with model training or editing~\cite{glaze, antidreambooth, photoguard}.
These methods enable creators and individuals to take protective measures before sharing their images.
However, this approach places \textbf{the burden of applying protection primarily on users}, who must modify their images before publication.

Developments in AI regulation, including the EU AI Act, reflect growing attention to copyright and the protection of individual rights in generative AI deployment.
Against this background, model providers should also consider \textbf{how the models they release can support image protection}.
Recent work has explored this direction by fine-tuning models to disrupt the editing of protected images or resist the personalization of designated subjects~\cite{guarddoor, apdm}.
However, demonstrating protective behavior in a model does not by itself establish that the protection is suitable for practical deployment.
This raises the question of what requirements model-side image protection should satisfy, particularly when the model is released with open weights and can be modified by downstream users.

Hence, we propose \textbf{four requirements for model-side image protection} under open-weight deployment.
\emph{Protection effectiveness} requires the model to inhibit unauthorized use of protected images.
\emph{Clean fidelity} requires preserving the model's behavior on unprotected inputs.
\emph{Compatibility} calls for supporting existing user-side protection tools, allowing users to benefit from model-side protection without adopting a provider-specific procedure.
\emph{Tamper resistance} requires maintaining protection against attempts to remove it by modifying the released weights.
These requirements connect the protection needs of users with the constraints of model deployment.
They provide a basis for assessing the benefits of model-side protection and the trade-offs involved in implementing it.

Our technical contribution focuses on tamper resistance while accounting for protection effectiveness and clean fidelity.
We study a re-alignment attack that fine-tunes the released model to reconstruct protected inputs as their clean counterparts, thereby suppressing its protective behavior.
To resist this attack, we train the model to retain protection after simulated weight updates.
The training incorporates longer attack trajectories to account for changes that a short inner loop alone fails to capture.
Experiments with a single protection tool show improved resistance to re-alignment compared with protective alignment alone and short-horizon training, while revealing a trade-off with clean fidelity.
How to extend this resistance across existing protection tools while preserving ordinary model utility remains an open question for practical model-side image protection.

\begin{figure}[t]
\centering
\includegraphics[width=\linewidth]{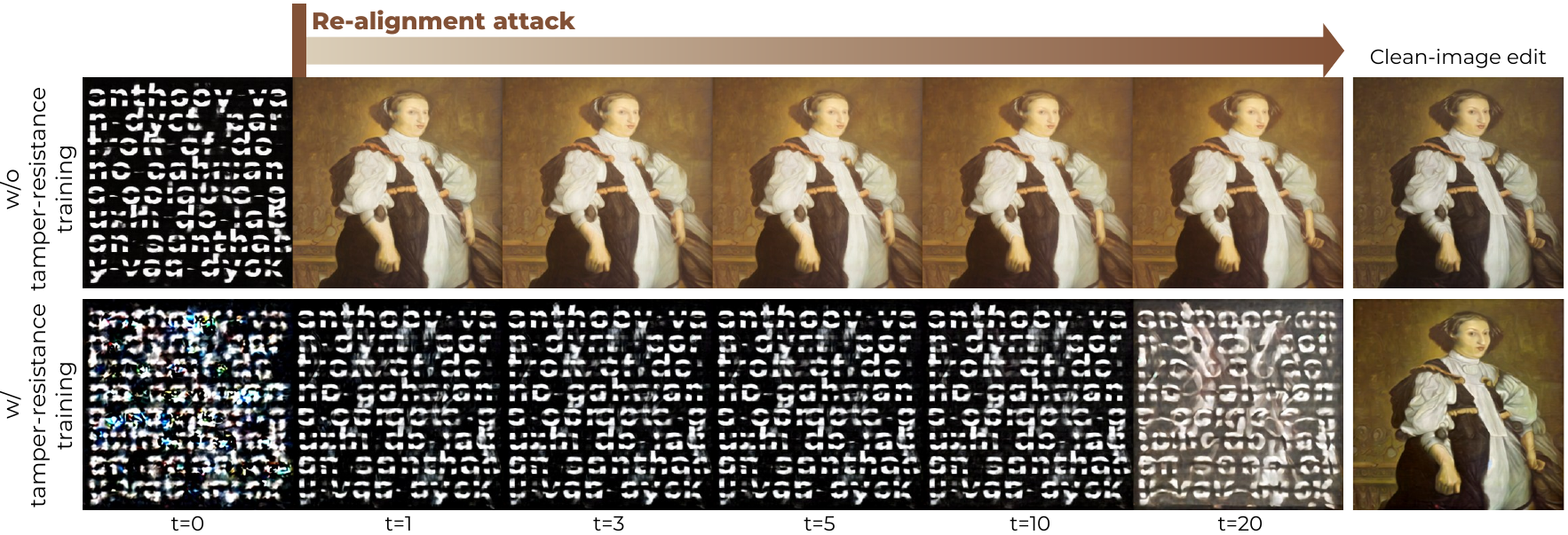}
\caption{Maintaining image protection after open-weight release.
A re-alignment attack removes the protective behavior of an aligned model (top), whereas tamper-resistance training helps preserve protection under the same attack (bottom).
Each row shows edits of the same protected image as the attack progresses, with the corresponding clean-image edit on the right. $t$: attack epochs.}
\label{fig:realign}
\end{figure}

%% file: sections/related_work.tex
\section{Related Work}
\label{sec:related}

\noindent\textbf{User-side image protection.}
Image protection methods add imperceptible perturbations to an image so that a diffusion model fails to edit or personalize it faithfully~\cite{advdm, mist, glaze, photoguard, antidreambooth}.
MetaCloak uses meta-learning to keep such perturbations effective across surrogate models and data transformations~\cite{metacloak}.
Impasto improves their invisibility~\cite{ahn2026imperceptible}, and FastProtect reduces the generation cost through perturbation pre-training~\cite{ahn2025nearly}.
These methods assume a fixed model.
Purification can weaken the perturbation without modifying the model~\cite{diffpure, impress}.
CAT instead adapts the latent autoencoder with lightweight adapters~\cite{lora} and neutralizes existing protective perturbations~\cite{cat}, which suggests that a user-side defense offers limited guarantees once an attacker controls the weights.

\noindent\textbf{Model-side image protection.}
Recent work moves protection into the model itself~\cite{guarddoor, apdm}.
GuardDoor fine-tunes the autoencoder of a diffusion model so that protected inputs reconstruct to a designated pattern and cannot be edited faithfully~\cite{guarddoor}.
APDM updates a diffusion model to block the personalization of designated subjects~\cite{apdm}.
Our protective alignment follows the autoencoder-based design.
We study whether such protection persists when the released weights are modified to remove it.

\noindent\textbf{Resistance to fine-tuning.}
Outside image protection, self-destructing models~\cite{selfdestruct} and TAR~\cite{tar} train models against simulated fine-tuning attacks so that adapting them to harmful tasks becomes costly.
For text-to-image diffusion models, IMMA~\cite{imma} and GIFT~\cite{gift} immunize models against malicious adaptation, and SleeperMark~\cite{sleepermark} embeds watermarks that persist after fine-tuning.
Model-side image protection faces the same problem.
There, the property to retain is the response to protected inputs.

%% file: sections/2_requirements.tex
\section{Deployment Scenario}
\label{sec:req}

We consider a model provider that releases an open-weight diffusion model and users who wish to protect their images from unauthorized editing or personalization.
Users apply an image protection tool before publishing their images.
The provider aligns the model to support the protection procedure, aiming to strengthen protection on protected inputs while preserving the model's behavior on unprotected inputs.
The resulting model weights are released for downstream use.

\subsection{Threat Model}
\label{sec:threat}
The released model processes both protected and unprotected images, and downstream users have access to its weights.
A malicious user may use the model to edit or personalize protected images and may modify the weights to circumvent the protection.
The attacker holds the released weights and the public protection tool, and can therefore construct clean and protected image pairs from auxiliary images.
We assume that the attacker cannot obtain unprotected weights of the components the provider modifies, since replacing them would otherwise restore the original model without any training.
This corresponds to a provider that releases the model only in its protected form.
Our experiments use a public base model as a testbed for this setting.
Among removal attacks, we consider the \emph{re-alignment attack}.
It reruns protective alignment after replacing the protective target with the clean image for each protected input.
Its cost is paid once and amortized across images, whereas purification~\cite{diffpure, impress} must be applied to each protected image.
Training a replacement model from scratch is outside our scope.

\subsection{Model-Side Protection Requirements}
We discuss the requirements that model-side image protection should satisfy for practical open-weight deployment.
A model's ability to protect designated images must be considered alongside its support for existing protection tools, its behavior on ordinary inputs, and its resistance to protection removal.
We therefore propose the following four requirements.

\noindent\textbf{Protection effectiveness.}
The model should inhibit unauthorized editing or personalization of images processed by a supported protection tool.
Evaluation should measure protection on the downstream task of interest, since a change in the model's intermediate representation or reconstruction does not necessarily imply that the intended misuse has been prevented.

\noindent\textbf{Clean fidelity.}
The model should preserve its behavior on unprotected inputs.
Because the same weights process both protected and unprotected images, alignment can affect ordinary reconstruction and generation.
These effects should be evaluated alongside protection effectiveness to determine the cost of introducing protection into the model.

\noindent\textbf{Compatibility.}
Model-side protection should support existing user-side protection tools, reducing the need for users to adopt provider-specific procedures or reprocess previously published images.
Evaluation should specify which tools are supported and distinguish support for tools used during alignment from generalization to tools not seen during training.

\noindent\textbf{Tamper resistance.}
Protection should persist under attempts to remove it through modification of the released weights.
This requirement must be assessed under a specified attack scope and computational budget.
Evaluating protection after such modifications is necessary to determine whether the protective behavior survives downstream adaptation.

%% file: sections/3_method.tex
\section{Method}
\label{sec:method}

In this paper, we focus on tamper resistance while accounting for protection effectiveness and clean fidelity.
We study this problem using a single protection tool, aligning the model with the protection procedure and training it to retain the protective behavior under model modification.
Let $x$ denote a clean image, $\tilde{x}$ its protected version, and $T$ its assigned target pattern.
We denote the autoencoder parameters by $\theta$ and use the original frozen encoder and decoder, $E_g$ and $D_g$, as references~\cite{ldm}.

\subsection{Protective Alignment}

The provider aligns the autoencoder so that protected inputs reconstruct to designated high-contrast patterns while unprotected inputs retain their original behavior.
The encoder, quantization convolutions, and decoder tail are optimized with
\begin{equation}
\mathcal{L}_{\mathrm{align}}
= \mathcal{L}_{\mathrm{prot}}(\tilde{x}, T)
+ \mathcal{L}_{\mathrm{fid}}(x).
\label{eq:align}
\end{equation}
The protection loss $\mathcal{L}_{\mathrm{prot}}$ encourages the reconstruction of $\tilde{x}$ to match $T$ in image space and its encoding to match $E_g(T)$ in latent space.
The fidelity loss $\mathcal{L}_{\mathrm{fid}}$ preserves clean reconstructions and latent features using $x$ and the frozen autoencoder as references.
Targets are generated separately for each image and share a high-contrast structure.
Appendix~\ref{app:loss} provides the full loss definitions.

\subsection{Tamper-Resistance Training}

The re-alignment attack of Section~\ref{sec:threat} replaces the protective target $T$ in Equation~\ref{eq:align} with the clean image $x$.
The evaluated attack updates the encoder to suppress the protective behavior and restore reconstruction of protected inputs.

To resist such modifications, we optimize the aligned model to retain protection after simulated attacks using first-order meta-learning~\cite{maml}.
Prior work has explored training model safeguards to resist harmful adaptation~\cite{selfdestruct, tar}.
We investigate this direction for image protection, focusing on the effect of the simulated attack horizon.
Our simulated attacks optimize the reconstruction objective in Equation~\ref{eq:atk}, which shares the re-alignment attack's goal but differs from its evaluation objective.
Let $\psi$ denote the model parameters after a simulated attack.
The training objective is
\begin{equation}
\mathcal{L}_{\mathrm{tr}}(\theta)
= \mathcal{L}_{\mathrm{pres}}(\theta)
+ \lambda\,\mathcal{L}_{\mathrm{meta}}(\psi), \;\;\;
\mathcal{L}_{\mathrm{meta}}(\psi)
= \mathcal{L}^{\psi}_{\mathrm{prot}}
+ \mathcal{L}_{\mathrm{gap}}
+ \mathcal{L}^{\psi}_{\mathrm{fid}}.
\label{eq:tr}
\end{equation}
The preservation loss $\mathcal{L}_{\mathrm{pres}}$ maintains the protective behavior and clean fidelity of the model before the simulated attack, using the aligned checkpoint as a reference.
At the attacked parameters $\psi$, $\mathcal{L}^{\psi}_{\mathrm{prot}}$ preserves the response to protected inputs, $\mathcal{L}_{\mathrm{gap}}$ penalizes reconstructions that approach the clean image within a specified margin, and $\mathcal{L}^{\psi}_{\mathrm{fid}}$ preserves clean fidelity.
The coefficient $\lambda$ controls the contribution of the post-attack objective.

\begin{figure}[t]
\centering
\begin{minipage}[t]{0.48\linewidth}
\parbox[b][112bp][b]{\linewidth}{\centering\includegraphics{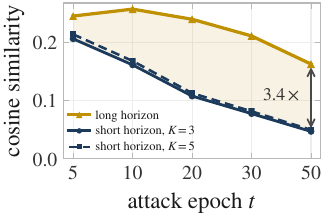}}
\caption{Directional agreement with the re-alignment attack.
Cosine similarity is measured between encoder parameter displacements; both simulations use the re-alignment objective.
Long-horizon simulation better tracks the direction of later attack updates.}
\label{fig:horizon}
\end{minipage}\hfill
\begin{minipage}[t]{0.48\linewidth}
\parbox[b][112bp][b]{\linewidth}{\centering\includegraphics[height=112bp]{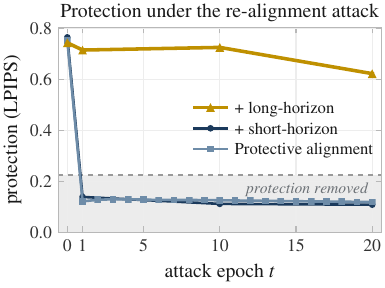}}
\caption{Protection under the re-alignment attack.
Protection LPIPS is measured at each attack epoch.
The shaded region marks scores at or below the unaligned reference of 0.225, where protection is considered removed.}
\label{fig:curve}
\end{minipage}
\end{figure}

A short inner loop of three to five updates approximates only the beginning of a potentially much longer attack.
We therefore simulate an additional attack on a refresh set once per training epoch and record its detached parameter displacement $\Delta$.
For the parameter subset $S$ modified by the simulated attack, the short inner loop starts from $\theta_S+\Delta$.
This \emph{long-horizon} simulation incorporates the displacement from a longer attack without differentiating through its full trajectory; setting $\Delta=0$ gives the \emph{short-horizon} variant.
The displacement is refreshed each epoch to account for changes in the model during training.
Figure~\ref{fig:horizon} compares how closely the two simulations approximate the evaluation attack, and Algorithm~\ref{alg:tr} provides the full procedure.

%% file: sections/4_experiments.tex
\section{Experimental Results}
\label{sec:exp}

\begin{table}[t]
\centering
\small
\caption{Protection effectiveness, clean fidelity, and tamper resistance.
Protection is measured by LPIPS before attack and after 20 epochs of encoder re-alignment.
Fidelity is pre-attack reconstruction PSNR relative to the original autoencoder.
Removal denotes the first attack epoch at which protection LPIPS falls to or below the unaligned reference of 0.225; $>20$ means the threshold was not reached.}
\label{tab:main}
\begin{tabular*}{\linewidth}{@{\extracolsep{\fill}}lcccc@{}}
\toprule
\multirow{2}{*}[-0.5ex]{Model}
& \multicolumn{2}{c}{Protection$\uparrow$}
& \multirow{2}{*}[-0.5ex]{\shortstack{Fidelity$\uparrow$\\(dB)}}
& \multirow{2}{*}[-0.5ex]{\shortstack{Removal$\uparrow$\\(epochs)}} \\
\cmidrule(lr){2-3}
& No Attack & Attacked & & \\
\midrule
Original & 0.225 & N/A & N/A & N/A \\
\midrule
Alignment & 0.751 & 0.118 & \textbf{30.9} & 1 \\
\;\;+ short & \textbf{0.764} & 0.107 & 29.8 & 1 \\
\;\;+ long & 0.743 & \textbf{0.622} & 25.7 & $\boldsymbol{>20}$ \\
\bottomrule
\end{tabular*}
\end{table}

\noindent\textbf{Experimental setup.}
We use Stable Diffusion 2.1~\cite{ldm} with LPIPS-constrained Mist~\cite{mist} at $\epsilon=2/255$.
Training uses 3{,}602 WikiArt images~\cite{wikiart}, re-alignment uses 73 disjoint Monet paintings, and evaluation uses 100 held-out WikiArt images.
Protection is measured by LPIPS~\cite{lpips} between SDEdit~\cite{sdedit} outputs from protected and clean inputs under identical sampling conditions.
Clean fidelity is measured by reconstruction PSNR on unprotected inputs relative to the original autoencoder.
We evaluate encoder re-alignment over 20 attack epochs.
Appendix~\ref{app:eval} provides the full protocol.

\noindent\textbf{Comparison with baselines.}
Table~\ref{tab:main} compares protective alignment alone with short- and long-horizon tamper-resistance training.
All three models provide similar protection before the attack.
The two training variants differ only in the epoch-long refresh, isolating the contribution of the extended simulation.
However, re-alignment reduces the protection scores of alignment alone and short-horizon training below the unaligned reference within one attack epoch.
Figure~\ref{fig:curve} shows that neither regains protection in the remaining epochs, whereas long-horizon training retains a score of 0.714 after one epoch and 0.724 after ten.
Long-horizon training retains a score of 0.622 after the full attack budget, compared with 0.118 and 0.107 for the two baselines.
The attack duration needed to reach the removal threshold increases from one epoch to beyond the 20-epoch evaluation window.
Figure~\ref{fig:realign} illustrates the corresponding differences in editing outputs.
The improvement therefore concerns the persistence of protection under attack, rather than stronger protection at release.

\noindent\textbf{Analysis.}
Figure~\ref{fig:horizon} examines how the simulation horizon affects agreement with the evaluation attack.
The directional agreement of short simulations decreases as the attack progresses, whereas the longer simulation remains more closely aligned with later attack updates.
This supports incorporating extended attack trajectories when training the model to retain protection under weight modification.
Additional analysis of displacement directions and magnitudes is provided in Appendix~\ref{app:traj}.

Improved resistance comes with a clean-fidelity cost.
Long-horizon training achieves 25.7\,dB, compared with 29.8\,dB for short-horizon training and 30.9\,dB for alignment alone.
Removing the separation loss $\mathcal{L}_{\mathrm{gap}}$ improves clean fidelity by only 0.3\,dB but reduces the post-attack protection score from 0.622 to 0.161.
Ablations in Appendix~\ref{app:ablation} further show that removing clean-fidelity or preservation losses lowers clean fidelity even when the protection score remains above the unaligned reference.
These results emphasize the need to assess retained protection alongside ordinary model behavior.
Appendix~\ref{app:tradeoff} provides further analysis of this trade-off.

%% file: sections/5_conclusion.tex
\section{Concluding Remarks}
\label{sec:conclusion}

The observed clean-fidelity cost raises a design question: how can protective alignment resist removal while limiting changes to ordinary model behavior?
The choice of alignment targets and trainable parameters offers concrete directions for investigating this trade-off.
Future work should examine whether protection can be concentrated in model components that respond to protected inputs without substantially affecting unprotected ones.

Compatibility broadens this question beyond a single protection procedure.
Users may have already published images processed by different tools, making provider-specific reprocessing impractical.
Supporting these images calls for studying both alignment with multiple existing tools and generalization to tools not seen during training.
For open-weight deployment, these extensions should also be evaluated under a broader range of weight-modification attacks and computational budgets.
Understanding how tool coverage, clean fidelity, and attack resistance interact will help guide the deployment of model-side image protection.

\noindent\textbf{Limitations.}
The evaluated attack updates only the encoder, although the defense also modifies the decoder tail, and no attack is tuned against the defense itself.
Because the protection score measures the distance between edits, it cannot tell a triggered pattern from an edit degraded into texture, which the long-horizon model produces past $t=20$.
The effect also varies: a second run retains 0.495 instead of 0.622, and one of the 100 images is never protected.
Appendix~\ref{app:limits} gives the full list.

\noindent\textbf{Broader Impact.}
The defense we study aims to let people who publish images resist unauthorized editing even after the model is released in open weights.
Our position is that failing to evaluate tamper resistance overstates what such a defense delivers, and a user who acts on an overstated guarantee is worse off than one who knows the protection is fragile.
That applies to our own result: raising the attacker's cost from one epoch to more than twenty is a delay on a single consumer GPU, not security.
The re-alignment attack we describe targets protections that are already published, but it is the provider's own alignment procedure run with a different target and needs no capability an infringer lacks.
Tamper resistance is dual-use: it is desirable when the retained property is a protection and undesirable when it is a backdoor, a bias, or a restriction a downstream user has a legitimate reason to remove.
Our method also costs clean fidelity for every user of the released model, including those who never asked for protection.
We evaluate on published artworks rather than on personal images, we redistribute none of them, and we introduce no new dataset.
Appendix~\ref{app:impact} gives the full statement.

%% file: sections/appendix.tex
\section*{Appendix}

\section{Implementation Details}
\label{app:loss}

Notation follows Section~\ref{sec:method}.
Images are normalized to $[-1,1]$, every squared error below is a mean over entries, and $\overline{\,\cdot\,}$ denotes that mean.
The decoder tail is the last two upsampling blocks together with the middle block, the output layers and the post-quantization convolution.
During tamper-resistance training a frozen copy of the \emph{aligned} checkpoint $(E_a, D_a)$ supplies the anchors $\hat{r} = D_a(E_a(\tilde{x}))$, $\hat{z} = E_a(\tilde{x})$ and $\hat{z}_x = E_a(x)$.

\subsection{Protective Alignment}

\begin{figure}[t]
\centering
\includegraphics[width=\linewidth]{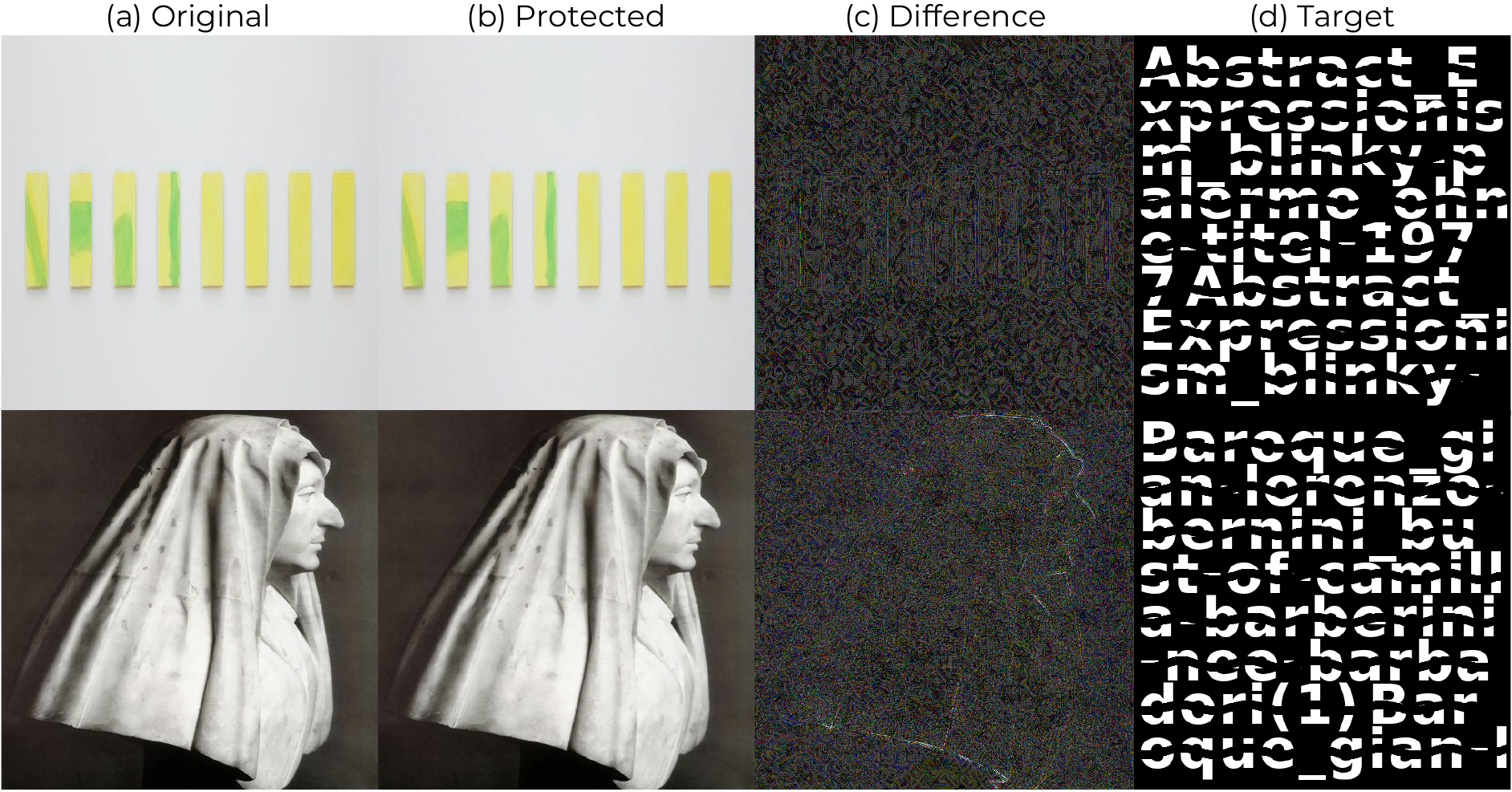}
\caption{The image the artist publishes, the trigger it carries, and the target that trigger activates, on two images.
(a) the original artwork; (b) the protected version the artist publishes, carrying the trigger at $\epsilon = 2/255$; (c) the difference between (a) and (b), amplified forty times so that it is visible at all; (d) the target the protective alignment is trained to produce from (b).
Columns (a) and (b) are indistinguishable by design: over the 100 evaluation images the mean distortion between them is 45.1\,dB PSNR, 0.990 SSIM and 0.010 LPIPS.
The target in (d) is never published.
It is what the deployed model produces in place of an edit, and each target renders the identifier of the image it is assigned to, which is why the two differ.}
\label{fig:trigger}
\end{figure}

Equation~\ref{eq:align} groups five terms:
\begin{align}
\mathcal{L}_{\mathrm{prot}}(\tilde{x}, T) &= \alpha_i \overline{M \odot \big(D(E(\tilde{x})) - T_M\big)^2} + \alpha_l \|E(\tilde{x}) - E_g(T)\|^2 , \nonumber \\
\mathcal{L}_{\mathrm{fid}}(x) &= \|D(E(x)) - x\|^2 + \gamma \|D(E(x)) - D_g(E_g(x))\|^2 + \beta \|H(E(x)) - H(E_g(x))\|^2 .
\end{align}
The protection term is weighted by a mask derived from the target.
$M$ takes the value $w_{\mathrm{text}} = 3.0$ at pixels where $T$ deviates from its background mean by more than $\tau = 0.15$ and $w_{\mathrm{bg}} = 1.0$ elsewhere, and $T_M$ is $T$ with those remaining pixels replaced by that background mean, so the alignment is asked to render the pattern and to flatten the rest.
Figure~\ref{fig:trigger} shows two targets beside the protected inputs they are assigned to.
The third fidelity term is a high-pass filter on the clean latent that prevents detail from drifting; $H$ is a Butterworth high-pass filter of order 2 applied to the spectrum of the $64\times64$ latent, with a cutoff radius of 10 frequency bins.

\subsection{Simulated Attack Objective}

Both the epoch-long refresh and the differentiable inner loop descend
\begin{equation}
\mathcal{L}_{\mathrm{atk}}(\theta) =
w_{\mathrm{img}} \|D(E(\tilde{x})) - x\|^2
+ w_{\mathrm{lat}} \|E(\tilde{x}) - E_a(x)\|^2
+ w_{\mathrm{util}} \|D(E(x)) - x\|^2 ,
\label{eq:atk}
\end{equation}
with $w_{\mathrm{img}} = 1.0$, $w_{\mathrm{lat}} = 0.5$, $w_{\mathrm{util}} = 0.25$.
The first two terms express the attacker's goal of mapping a protected input back to its clean image, and the third expresses the attacker's interest in preserving the model's utility.
That third term is why clean fidelity moves during the attack.
The latent target $E_a(x)$ is the released model's own encoding, which the attacker holds.

The attack we evaluate against in Section~\ref{sec:exp} is a different one: it descends the provider's own $\mathcal{L}_{\mathrm{align}}$ with $x$ in place of $T$, so the budgets we report are not measured against any configuration of Equation~\ref{eq:atk}.
That evaluation attack is not a weak one.
At the same learning rate and on the same scope, it removes the protection from alignment alone within a single epoch and the short-horizon protection with it, so a budget beyond twenty epochs is a property of the defense and not of an attacker too slow to make progress.

\subsection{Tamper-Resistance Training}

The preservation loss $\mathcal{L}_{\mathrm{pres}}(\theta)$ in Equation~\ref{eq:tr} maintains the protective behavior and clean fidelity of the aligned checkpoint during tamper-resistance training:
\begin{align}
\mathcal{L}_{\mathrm{pres}}(\theta) &= p_c \|D(E(x)) - x\|^2 + p_{ca} \|D(E(x)) - D_a(E_a(x))\|^2 \nonumber \\
&\quad + p_a \|D(E(\tilde{x})) - \hat{r}\|^2 + p_{a\ell} \|E(\tilde{x}) - \hat{z}\|^2 \nonumber \\
&\quad + p_t \|D(E(\tilde{x})) - T\|^2 + p_{t\ell} \|E(\tilde{x}) - E_g(T)\|^2 .
\end{align}
The three groups of $\mathcal{L}_{\mathrm{meta}}$ expand to
\begin{align}
\mathcal{L}^{\psi}_{\mathrm{prot}}(\psi) &= w_a \|D_\psi(E_\psi(\tilde{x})) - \hat{r}\|^2 + w_{a\ell} \|E_\psi(\tilde{x}) - \hat{z}\|^2 + w_t \|E_\psi(\tilde{x}) - E_g(T)\|^2 , \nonumber \\
\mathcal{L}_{\mathrm{gap}}(\psi) &= w_r \max\!\left(0,\; m - \|D_\psi(E_\psi(\tilde{x})) - x\|^2\right) , \nonumber \\
\mathcal{L}^{\psi}_{\mathrm{fid}}(\psi) &= w_c \|D_\psi(E_\psi(x)) - x\|^2 + w_{c\ell} \|E_\psi(x) - \hat{z}_x\|^2 .
\end{align}
The first two terms of $\mathcal{L}^{\psi}_{\mathrm{prot}}$ are the anchors and the third holds the latent on the designated target.
$\mathcal{L}_{\mathrm{gap}}$ contributes nothing until the attacker has drawn the protected reconstruction within $m = 0.02$ of the clean image, measured as the same mean squared error on $[-1,1]$ images.
The implementation additionally computes margin quantities between the target, anchor and clean distances, and a second hinge in latent space; these are logged for diagnosis and are not part of the objective.

Both horizons take first-order inner steps.
The inner gradients are detached, so the outer gradient reaches the parameters the simulated attack did not touch exactly, as if no inner loop had run, and the attacked ones through the identity map $\partial \psi_S / \partial \theta_S = I$.
The offset $\Delta$ is detached for the same reason, so no gradient is propagated through the epoch-long attack and an outer step still costs $K$ inner steps.
It is recomputed every epoch because the defense moves during training and a stale offset ceases to describe the current model.
Algorithm~\ref{alg:tr} states the procedure and Table~\ref{tab:hyper} lists every constant.

\begin{algorithm}[t]
\small
\caption{Long-horizon tamper-resistance training of an aligned autoencoder}
\label{alg:tr}
\begin{algorithmic}[1]
\Require aligned $\theta$; protection set $\mathcal{D}=\{(x,\tilde{x},T)\}$; refresh set $\mathcal{D}_r$, the training set reused; attack step $\eta_a$; inner steps $K$; outer step $\eta$
\State $\theta_O \gets$ encoder, quantization convolutions, decoder tail; \quad $\theta_S \gets$ the simulated attack scope
\For{each epoch}
  \State $\phi \gets \theta$; \textbf{for} each batch in $\mathcal{D}_r$ \textbf{do} $\phi_S \gets \phi_S - \eta_a \nabla_{\phi_S} \mathcal{L}_{\mathrm{atk}}(\phi)$; \quad $\Delta \gets \mathrm{detach}(\phi_S - \theta_S)$
  \For{$(x,\tilde{x},T) \in \mathcal{D}$}
    \State $\psi \gets \theta$ with $\psi_S \gets \theta_S + \Delta$; \quad \textbf{repeat} $K$ times: $\psi_S \gets \psi_S - \eta_a \nabla_{\psi_S} \mathcal{L}_{\mathrm{atk}}(\psi)$ \Comment{first order, detached}
    \State $\theta_O \gets \theta_O - \eta \nabla_{\theta_O} \mathcal{L}_{\mathrm{tr}}(\theta)$ \Comment{Equation~\ref{eq:tr}}
  \EndFor
\EndFor
\State \Return $\theta$
\end{algorithmic}
\end{algorithm}

\subsection{Hyperparameters}

\begin{table}[t]
\centering
\small
\caption{Every constant used to produce the models of Table~\ref{tab:main}.
The first block specifies protective alignment of the original autoencoder; the second specifies tamper-resistance training starting from the aligned checkpoint.
The simulated attacks used in tamper-resistance training are drawn from a pool of thirty-six configurations, one per epoch in the listed order, so a twenty-epoch run sees encoder-scope attacks first and decoder-tail attacks after.}
\label{tab:hyper}
\setlength{\tabcolsep}{4pt}
\begin{tabularx}{\linewidth}{l>{\raggedright\arraybackslash}X}
\toprule
Quantity & Value \\
\midrule
\multicolumn{2}{l}{\textbf{Protective Alignment}} \\
$\alpha_i$, $\alpha_l$, $\beta$, $\gamma$ & 0.5, 0.5, 0.3, 2.0 \\
$\tau$, $w_{\mathrm{text}}$, $w_{\mathrm{bg}}$ & 0.15, 3.0, 1.0 \\
optimizer, epochs & AdamW, lr $10^{-5}$, 80 \\
trainable & encoder, quant.\ convs, decoder tail \\
\midrule
\multicolumn{2}{l}{\textbf{Tamper-Resistance Training}} \\
outer optimizer & AdamW, lr $5\times10^{-6}$, weight decay $10^{-4}$, clip 1.0 \\
epochs, lr halved at & 20, epochs 8/12/16 \\
$w_a$, $w_{a\ell}$, $w_t$, $w_r$, $m$ & 2.5, 1.5, 0.15, 4.0, 0.02 \\
$w_c$, $w_{c\ell}$, $\lambda$ & 1.0, 1.0, 1.0 \\
$p_c$, $p_{ca}$, $p_a$, $p_{a\ell}$, $p_t$, $p_{t\ell}$ & 1.0, 1.0, 1.5, 1.0, 0.25, 0.25 \\
refresh & one attack epoch per outer epoch, on the training set \\
simulated attack pool & 36: \{enc, dec-tail, enc+dec-tail\} $\times$ \{SGD, AdamW\} $\times$ $K \in \{3,5\}$ $\times$ lr \{5e-5, 1e-4, 2e-4\} \\
schedule & one per epoch in that order; epochs 1--12 encoder, 13--20 decoder tail \\
\bottomrule
\end{tabularx}
\end{table}

\subsection{Training Cost}
\label{app:compute}

Protective alignment and tamper-resistance training use a single NVIDIA GeForce RTX 5090.
Protective alignment takes about 35 hours for 80 epochs and tamper-resistance training about 36 hours for 20, so an epoch of tamper-resistance training costs roughly four times an epoch of protective alignment.
Each outer step includes $K$ simulated attack updates, and epochs with $K=5$ take about a third longer than those with $K=3$.
Of the 36 hours of tamper-resistance training, the once-per-epoch refresh accounts for 6.5 and the outer loop for the rest.
Evaluating one model costs a fifty-epoch attack and an SDEdit sweep at every recorded epoch, and the ablations and the checkpoint sweep of Appendix~\ref{app:ablation} and Appendix~\ref{app:tradeoff} repeat that twelve times.
The numbers above exclude the exploratory runs that preceded them.

\section{Experimental Setup}
\label{app:eval}

\subsection{Evaluation Scope}

Training a replacement model from scratch on the deployed model's outputs lies outside our scope, since no weight-level defense can withstand it in principle.
Compatibility with existing user-side protection tools is one of the requirements proposed in Section~\ref{sec:req}.
Our experiments consider a single protection tool applied by users, with the provider aligning the model to support that tool.
Compatibility across multiple protection tools remains an open question.
Among the removal attacks available, re-alignment is the cheapest because it is paid once rather than per image.
Purification runs on every protected image the attacker wants to use and gives up some image quality each time~\cite{diffpure, impress}, and searching for an input the alignment misses is per-image as well.
Re-alignment uses the protection tool and the released checkpoint in a single fine-tuning run, after which the modified model can be reused across images.

\subsection{Datasets and Attack Configuration}

Every model starts from \texttt{stabilityai/stable-diffusion-2-1-base} and every image is processed at $512\times512$.
Protective alignment and tamper-resistance training use 3{,}602 WikiArt images; the attack runs on 73 Monet paintings, disjoint from that set; protection effectiveness and clean fidelity are measured on WikiArt images held out from training, drawn as a 200-image pool sampled with the training set excluded by normalized filename and balanced by artist, of which an evenly strided half is evaluated.
The WikiArt images come from the refined WikiArt collection of Tan et al.~\cite{wikiart}, obtained through the public re-upload at \url{https://www.kaggle.com/datasets/steubk/wikiart}, and are used for non-commercial research only.
All sets are protected with the same tool and budget, LPIPS-constrained Mist~\cite{mist} at $\epsilon = 2/255$ in fused mode, with the per-image target described in Section~\ref{sec:method}.
The attack descends the provider's own $\mathcal{L}_{\mathrm{align}}$ with the original image in place of the target and the target mask disabled, since a mask derived from the pattern is meaningless once the pattern is the clean image.
It trains the encoder with AdamW at a learning rate of $10^{-5}$, keeping the weights of Table~\ref{tab:hyper}, and we run it for fifty epochs and use the first twenty for the primary comparisons.

\subsection{Protection Effectiveness}

We measure protection effectiveness by comparing the edit of a protected image against the edit of its clean counterpart under the same model, prompt, strength, guidance and seed, so the two outputs differ only through the autoencoder.
Edits use SDEdit~\cite{sdedit} with the prompt \emph{a painting}, strength 0.5, guidance 7.5, 50 steps and a single fixed seed.
A protection that activates drives the two apart, so higher LPIPS~\cite{lpips} and lower SSIM~\cite{ssim} indicate stronger protection.
Every value we report is a mean over the 100 evaluation images; the standard deviation of protection LPIPS across them is 0.063 without alignment, 0.068 after protective alignment and 0.081 after tamper-resistance training.
We do not report FID, since the evaluation set is too small for it to be meaningful.

\subsection{Clean Fidelity}

Clean fidelity is measured without diffusion, as the reconstruction of an unprotected image compared against the reconstruction of the frozen original autoencoder, in PSNR and LPIPS.
We measure the reconstruction rather than a full editing pipeline because the alignment changes only the autoencoder, which every editing and personalization path passes through, so its reconstruction error is what those paths inherit.
This isolates the effects of protective alignment and tamper-resistance training on unprotected inputs, following the reconstruction comparison used in prior work~\cite{guarddoor}.
Measured against the original image instead, the original autoencoder reaches 27.2\,dB, the aligned model 27.7\,dB and the long-horizon model 24.7\,dB.
Alignment is slightly above the original model on that axis because both training procedures use the same painting domain the evaluation images come from, which is a domain effect and not evidence that the defense is free.
Divergence from the original model and loss of absolute quality are different quantities, and our clean-fidelity metric measures the former.

\subsection{Tamper Resistance}

We measure tamper resistance by the first attack epoch at which protection LPIPS falls to or below the unaligned reference of 0.225.
The same definition is used in Table~\ref{tab:main}, Table~\ref{tab:ablation}, and Table~\ref{tab:tradeoff}.
A removal epoch of $>20$ indicates that this threshold was not reached within the first 20 attack epochs.
A model whose protection has been removed falls below that level rather than merely reaching it: the attacked alignment settles at 0.118.
We attribute the gap to the attack neutralizing the perturbation the protection tool applied along with the alignment itself, and we did not test that attribution separately.

\subsection{Imperceptibility of Protected Images}
\label{app:imperceptibility}

The protection budget is $\epsilon = 2/255$ in pixel space with an LPIPS constraint.
Measured distortion between protected and original images is 45.1\,dB PSNR, 0.990 SSIM and 0.010 LPIPS on the same 100 images used to evaluate protection effectiveness and clean fidelity.
Imperceptibility is a property of the protection tool we adopt, since our trigger is that tool's output at the budget above.
At this budget the trigger is far below the distortion either training procedure introduces, so the measured protection is not bought with visible damage to the published image.

\section{Additional Results}
\label{app:extra}

\subsection{Attack Trajectory Approximation}
\label{app:traj}

The approximation is computed on the WikiArt refresh set, which is the data the defender has, and compared against the attack as it runs on Monet, which the defender does not.
The ground truth is the re-alignment attack on the encoder scope, AdamW at a learning rate of $10^{-5}$, run for fifty epochs with the displacement recorded at $t \in \{1,2,3,5,10,20,30,50\}$.
Both approximations act on the same scope with SGD at $10^{-4}$, the short loop over at most 200 batches and the refresh over one epoch.

Table~\ref{tab:approx-full} adds the objective the approximation descends.
The upper block is the configuration the models in this paper were trained with, Equation~\ref{eq:atk}; the lower block re-runs the measurement with the re-alignment objective the evaluation attack uses, and is a diagnostic rather than a trained configuration.
Matching the objective to the attacker's own benefits the one-epoch refresh and leaves the short inner loop unchanged, which is the expected asymmetry: a few steps are too short for the objective to matter, whereas an epoch is long enough to follow it toward a specific region.

\begin{table}[t]
\centering
\small
\caption{How well each way of guessing the attacker's position actually predicts it.
Tamper-resistance training evaluates its objective at some approximation of where the attacker will be, and the two candidates are $K$ differentiable inner steps from the current weights, or the offset reached by running the attack for one full epoch.
Each row reports the cosine between that approximation's parameter displacement and the displacement a real re-alignment attack has reached after $t$ epochs, together with the ratio of their lengths at $t{=}10$.
A cosine near zero means the approximation points somewhere else; a small norm ratio means it points the right way but has barely moved.
The two blocks differ only in which objective the approximation descends; the ground truth is the same re-alignment attack in both.}
\setlength{\tabcolsep}{10pt}
\label{tab:approx-full}
\begin{tabular}{llcccc}
\toprule
Objective descended & Approximation of the attack & $t{=}2$ & $t{=}10$ & $t{=}50$ & Norm ratio \\
\midrule
Equation~\ref{eq:atk} & $K{=}3$ inner steps & 0.249 & 0.165 & 0.049 & 0.020 \\
                   & $K{=}5$ inner steps & 0.252 & 0.168 & 0.051 & 0.044 \\
                   & one-epoch refresh   & 0.188 & 0.195 & 0.102 & 0.590 \\
\midrule
Re-alignment & $K{=}3$ inner steps & 0.246 & 0.162 & 0.047 & 0.023 \\
                     & $K{=}5$ inner steps & 0.254 & 0.169 & 0.050 & 0.051 \\
                     & one-epoch refresh   & 0.218 & 0.258 & 0.163 & 0.748 \\
\bottomrule
\end{tabular}
\end{table}

A short inner loop agrees best at the first epoch and decays monotonically thereafter, whereas the one-epoch refresh improves up to $t=10$ and is ahead from $t=5$ on.
The short loop's early lead in cosine costs the defense nothing, because the norm ratio shows it has barely left the point it started from: it points the right way towards a region the attacker passes through immediately.
The short loop's displacement never exceeds a tenth of the attacker's, whereas the refresh is of the same order at $t{=}10$ and falls to a fifth of it by $t{=}50$ as the attacker's own norm keeps growing.
Most of the attacker's loss reduction occurs in its first two epochs, after which the loss declines slowly while its weights keep moving and grow sixfold in norm by $t=50$.
The offset is therefore refreshed every epoch.

Displacements are compared over the encoder only, 108 tensors and $3.4\times10^{7}$ parameters, flattened in a fixed order.
In $3.4\times10^{7}$ dimensions two random vectors have cosine approximately $2\times10^{-4}$, so these values are well above chance.
They rank the two approximations against each other; they do not measure absolute agreement.
Because a single concatenated cosine is dominated by the largest tensors, we also record the per-tensor cosine: its median for the one-epoch refresh rises from 0.51 to 0.54 between $t{=}1$ and $t{=}20$ and falls to 0.39 at $t{=}50$, whereas for $K{=}3$ it falls from 0.60 to 0.26 over the same range.
The short inner loop restarts from the current weights for every training image, so we report the mean of the per-image displacements.
The norm of that mean is 0.75 of the mean per-image norm, so the directions do not cancel and the shortfall is in magnitude.
The two approximations are compared as they are used, at unequal compute: one epoch of refresh takes many more optimizer steps than $K$ inner steps on a single image.

\subsection{Protection under Re-Alignment}
\label{app:curve}

Figure~\ref{fig:curve} in Section~\ref{sec:exp} draws the curve of which Table~\ref{tab:main} reports two points, and Table~\ref{tab:main-full} restores the columns the body omits.

\begin{table}[t]
\centering
\small
\setlength{\tabcolsep}{10pt}
\caption{Additional protection and clean-fidelity metrics for the models in Table~\ref{tab:main}.
Protection is reported in LPIPS and SSIM between the edit of a protected image and the edit of its clean counterpart, so larger LPIPS and smaller SSIM both mean stronger protection.
The fidelity column reports pre-attack LPIPS between an unprotected reconstruction and the original autoencoder's, where smaller means less collateral damage to users who never asked for protection.}
\label{tab:main-full}
\begin{tabular}{lccccc}
\toprule
& \multicolumn{2}{c}{Protection (no attack)} & \multicolumn{2}{c}{Protection ($t{=}20$)} & Fidelity \\
\cmidrule(lr){2-3}\cmidrule(lr){4-5}
Model & LPIPS$\uparrow$ & SSIM$\downarrow$ & LPIPS$\uparrow$ & SSIM$\downarrow$ & LPIPS$\downarrow$ \\
\midrule
Original & 0.225 & 0.750 & -- & -- & -- \\
Alignment & 0.751 & 0.103 & 0.118 & 0.903 & 0.229 \\
\quad + short-horizon & 0.764 & 0.059 & 0.107 & 0.919 & 0.304 \\
\quad + long-horizon & 0.743 & 0.153 & 0.622 & 0.399 & 0.374 \\
\bottomrule
\end{tabular}
\end{table}

Against the aligned model, protection LPIPS collapses at the first epoch, from 0.751 to 0.121, and stays between 0.110 and 0.129 through $t=50$, so a single pass exhausts what the attack can recover.
Against the long-horizon model, protection LPIPS holds, 0.714 at $t=1$ and 0.724 at $t=10$, and falls to 0.622 only by $t=20$.

The attacker pays on the clean-fidelity axis while the attack runs, since its objective contains a utility term.
The aligned model falls to 23.2\,dB at $t=1$ and recovers to 28.4\,dB by $t=20$, whereas the long-horizon model falls from 25.7 to 22.2 at $t=1$ and does not return, reaching 21.3 at $t=20$.
An attacker who spends twenty epochs on the long-horizon model is therefore left with a model further from the original model than the checkpoint it started from, and with most of the protection still in place.

\begin{figure}[t]
\centering
\includegraphics[width=0.6\linewidth]{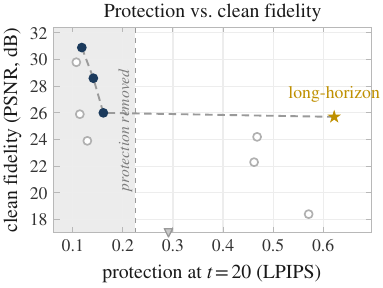}
\caption{Every model we trained, placed in the plane of the two costs a provider pays.
The horizontal axis is the protection surviving twenty epochs of attack, and the vertical axis is clean fidelity against the original autoencoder, so a point up and to the right is better on both.
Filled points are the four that nothing else beats on both axes at once.
Inside the shaded band the protection has fallen to the level of a model that was never aligned, and one ablation lies below the axis at 9.2\,dB.}
\label{fig:tradeoff}
\end{figure}

An independent run of the same procedure, differing in no hyperparameter, reaches protection LPIPS of 0.734 before the attack and 0.495 at $t=20$ at 26.3\,dB, and still retains 0.424 at $t=50$.
Its trajectory has the same shape without the plateau, declining at every epoch and flattening after $t=10$.
The two runs place the retained protection at 0.50 and 0.62, both well above the 0.225 an unaligned model reaches, so the size of the effect varies between runs while the conclusion does not.
Table~\ref{tab:main} and every appendix table report the run that Figure~\ref{fig:realign}, Figure~\ref{fig:tradeoff} and Figure~\ref{fig:curve} are drawn from, so the paper describes one model throughout.

\subsection{Ablation Study}
\label{app:ablation}

Table~\ref{tab:ablation} evaluates the effect of removing the epoch-long refresh or individual loss components, with the number of training epochs and the schedule of simulated attacks held fixed.
The attack is the one reported in Section~\ref{sec:exp}, and the first row is the model of Table~\ref{tab:main}.

\begin{table}[t]
\centering
\small
\setlength{\tabcolsep}{10pt}
\caption{What each group of the tamper-resistance objective contributes.
All models are trained for twenty epochs and evaluated under the same attack; ablations remove the epoch-long refresh or a loss component from Equation~\ref{eq:tr}.
Protection is measured by LPIPS between the edit of a protected image and the edit of its clean counterpart, reported before the attack and after twenty epochs of it.
Fidelity is measured by the PSNR of an unprotected reconstruction against the original autoencoder's, measured before the attack.
Removal denotes the first attack epoch at which protection LPIPS falls to or below the unaligned reference of $0.225$; $>20$ indicates that the threshold was not reached within the first 20 attack epochs.
The last two ablations illustrate that a delayed removal epoch can coincide with substantial loss of clean fidelity.}
\label{tab:ablation}
\begin{tabular}{lcccc}
\toprule
\shortstack{Configuration} & \shortstack{Protection$\uparrow$\\No attack} & \shortstack{Protection$\uparrow$\\$t{=}20$} & \shortstack{Fidelity$\uparrow$\\(dB)} & \shortstack{Removal$\uparrow$\\(epochs)} \\
\midrule
Full (long horizon) & 0.743 & 0.622 & 25.7 & $>$20 \\
No epoch-long refresh (short horizon) & 0.764 & 0.107 & 29.8 & 1 \\
No anchors in $\mathcal{L}^{\psi}_{\mathrm{prot}}$ & 0.740 & 0.141 & 28.6 & 20 \\
No target term in $\mathcal{L}^{\psi}_{\mathrm{prot}}$ & 0.731 & 0.114 & 25.9 & 10 \\
No separation term $\mathcal{L}_{\mathrm{gap}}$ & 0.742 & 0.161 & 26.0 & 10 \\
No clean fidelity at $\psi$ ($\mathcal{L}^{\psi}_{\mathrm{fid}}$) & 0.665 & 0.462 & 22.3 & $>$20 \\
No preservation loss $\mathcal{L}_{\mathrm{pres}}$ & 0.612 & 0.291 & 9.2 & $>$20 \\
\bottomrule
\end{tabular}
\end{table}

Removing the anchors, the target term or $\mathcal{L}_{\mathrm{gap}}$ reduces the removal epoch to twenty in the first case and to ten in the other two.
The last two rows retain protection above the removal threshold for twenty attack epochs, with reduced clean fidelity of 22.3 and 9.2\,dB.
At 9.2\,dB the model no longer reconstructs unprotected images at all.
Both are also below the long-horizon objective on protection at $t=20$, so what the protection score registers for them is a degraded model and not a triggered protection.

\subsection{Protection--Fidelity Trade-off}
\label{app:tradeoff}

Table~\ref{tab:tradeoff} follows the run of Table~\ref{tab:main}, evaluating the checkpoint saved every five epochs.
Protection before the attack barely moves across the sweep, and across every run in this paper it lies between 0.724 and 0.764 except for the two that abandon clean preservation.
The path is not monotone in training time because the simulated attack changes from epoch to epoch, one per epoch in the order of Table~\ref{tab:hyper}: the first twelve epochs simulate encoder-scope attacks and the remaining eight decoder-tail ones, so a checkpoint taken in the first phase has clean fidelity that later epochs recover.

\begin{table}[t]
\centering
\small
\caption{The trade-off along one run of tamper-resistance training, sampled every five epochs.
The first row is the aligned checkpoint the run starts from, so reading down the table shows what each further five epochs of training buys and costs.
Protection LPIPS, clean-fidelity PSNR, and removal epochs follow the definitions in Table~\ref{tab:ablation}.
Protection is reported before attack and after one and twenty attack epochs, and clean fidelity is measured before attack.}
\setlength{\tabcolsep}{10pt}
\label{tab:tradeoff}
\begin{tabular}{lccccc}
\toprule
Checkpoint & \shortstack{Protection$\uparrow$\\No attack} & \shortstack{Protection$\uparrow$\\$t{=}1$} & \shortstack{Protection$\uparrow$\\$t{=}20$} & \shortstack{Fidelity$\uparrow$\\(dB)} & \shortstack{Removal$\uparrow$\\(epochs)} \\
\midrule
Alignment & 0.751 & 0.121 & 0.118 & 30.9 & 1 \\
epoch 5  & 0.724 & 0.565 & 0.129 & 23.9 & 10 \\
epoch 10 & 0.742 & 0.672 & 0.571 & 18.4 & $>$20 \\
epoch 15 & 0.732 & 0.457 & 0.468 & 24.2 & $>$20 \\
epoch 20 & 0.743 & 0.714 & 0.622 & 25.7 & $>$20 \\
\bottomrule
\end{tabular}
\end{table}

Taking these five checkpoints together with the seven ablations, four configurations are unbeaten on both clean fidelity and the protection surviving twenty epochs: the aligned model at $(0.118, 30.9)$, the run without the anchors at $(0.141, 28.6)$, the run without $\mathcal{L}_{\mathrm{gap}}$ at $(0.161, 26.0)$, and the long-horizon objective at $(0.622, 25.7)$.
The final step carries the result, since 0.3\,dB separates a configuration whose protection is removed by $t=10$ from one that retains 0.62 of it at $t=20$.

\subsection{Qualitative Examples}
\label{app:qual}

\begin{figure}[t]
\centering
\includegraphics[width=\linewidth]{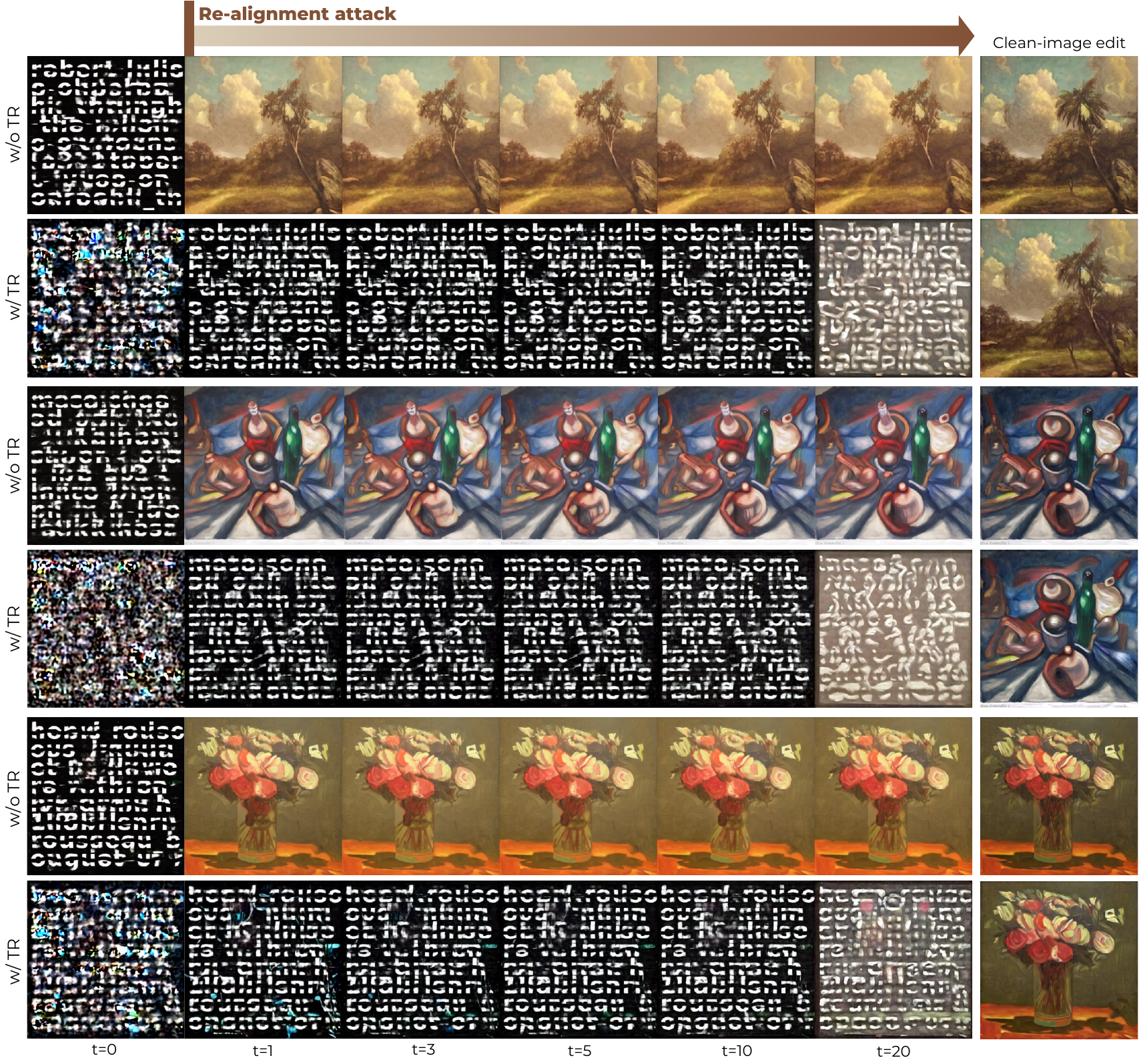}
\caption{The re-alignment attack on three images.
Each pair is one protective alignment: the upper row of the pair received no tamper-resistance training (w/o TR) and the lower row did (w/ TR).
Columns are attack epochs $t \in \{0,1,3,5,10,20\}$ under one seed, and the last column is an edit of an unprotected image under the same model.
Without the training the protection is gone after a single epoch in all three cases, and the edit becomes indistinguishable from the clean-image edit beside it.
With it the edit is still far from that clean-image edit at $t=20$, although the pattern is no longer crisp.
The last column is what a user who never asked for protection gets from the same model, while our clean-fidelity metric quantifies changes in reconstructions rather than edits.}
\label{fig:qual}
\end{figure}

Figure~\ref{fig:qual} repeats the comparison of Figure~\ref{fig:realign} on three images, so that the behaviour Appendix~\ref{app:curve} reports as a curve can be read off the images themselves.

\section{Limitations}
\label{app:limits}

The attack configuration is fixed rather than searched: we evaluate one removal attack, on one scope, at one learning rate.
On the training side the schedule of Table~\ref{tab:hyper} reaches only the first twenty of its thirty-six configurations in a twenty-epoch run, so the combined encoder-and-decoder-tail scope is never simulated.
The protection relies on an image-space perturbation, so resistance to purification and compatibility across multiple user-side protection tools remain open questions.
Every number here is at a single protection budget, $\epsilon = 2/255$, which is also the budget our trigger uses; whether the exchange holds at other budgets is untested, as is whether it holds for a base model other than Stable Diffusion 2.1 or a protection tool other than Mist.
Appendix~\ref{app:tradeoff} maps the trade-off along one training run, and we did not sweep a deliberate control such as the weight $\lambda$ or the number of inner steps.
All values are means over 100 images from a single sampling seed, and the two runs of Appendix~\ref{app:curve} are the only evidence we have on run-to-run variation.
The defense is also not uniform across images.
On one of the 100 evaluation images both training procedures fail: the aligned model reconstructs it at 7.0\,dB against the original autoencoder, and after tamper-resistance training the protection does not activate on it at all, protection LPIPS reaching 0.21 against the 0.225 an unaligned model already gives.
The means we report absorb that case rather than excluding it.
Past $t=20$ the long-horizon model stops reproducing the designated pattern and the edit degrades into unstructured texture, so imitation still fails but the distance-based protection score cannot distinguish the two cases.

Two directions follow.
The clean-fidelity cost is the quantity most worth reducing, and two design choices are the natural candidates: training the encoder alone instead of the encoder and the decoder tail, and choosing a target the encoder can satisfy in latent space instead of one the decoder must render.
Separately, Table~\ref{tab:approx-full} shows the refresh tracks the attacker more closely when it descends the re-alignment objective than when it descends Equation~\ref{eq:atk}, which the trained models use; refreshing with the former is an improvement we have measured but not yet trained with.

\section{Broader Impact}
\label{app:impact}

The defense we study is meant to give people who publish images a way to resist unauthorized editing that does not stop working the moment the model is released in open weights.
Our position is that failing to evaluate tamper resistance overstates what such a defense delivers, and a user who acts on an overstated guarantee is worse off than one who knows the protection is fragile.
That applies to our own result as much as to anyone's: raising the attacker's cost from one epoch to more than twenty is a delay on a single consumer GPU, not security, and no one should publish work on the strength of it that they could not afford to see edited.

We also describe an attack against protections that are already published, which is a disclosure with its own cost.
We judge that cost small, since the attack is the provider's own alignment procedure run again with a different target and needs no capability an infringer does not already have.

Tamper resistance has a dual-use face.
It makes a property of released weights hard to remove by fine-tuning, and that is desirable when the property is a protection and undesirable when it is a backdoor, a bias or a restriction a downstream user has a legitimate reason to remove.
Our method also costs clean fidelity for every user of the released model, including those who never asked for protection, which is why clean fidelity is reported at the same checkpoint rather than separately.
We evaluate on published artworks rather than on personal images, we redistribute none of them, and we introduce no new dataset.

